\documentclass[final,5p,times,twocolumn]{article}

\usepackage[utf8]{inputenc}
\usepackage{natbib}
\usepackage{graphicx}
\usepackage[margin=0.5in]{geometry}

\usepackage{lineno,hyperref}
\modulolinenumbers[5]

\usepackage{epstopdf}
\usepackage{verbatim}
\usepackage{mathrsfs}
\usepackage{array}

\newcolumntype{H}{>{\setbox0=\hbox\bgroup}c<{\egroup}@{}}

\begin{document}

\title{Asymmetry in the distribution of HSC galaxy spin directions: comment on arXiv: 2410.18884v1}


\date{}

\author{Lior Shamir\footnote{lshamir@mtu.edu}  \\ Kansas State University \\ Manhattan, KS 66506, USA}

\maketitle

\begin{abstract}
In the past decade, an asymmetry in the large-scale distribution of galaxy spin directions has been observed in data from all relevant digital sky surveys, all showing a higher number of galaxies rotating in the opposite direction relative to the Milky Way as observed from Earth. Additionally, JWST deep fields have shown that the asymmetry is clear and obvious, and can be sensed even by the naked human eye. These experiments were performed using two separate statistical methods: standard binomial distribution and simple $\chi^2$ statistics. Stiskalek \& Desmond (2024) suggested that the asymmetry in the distribution of galaxy spin directions is due to the use of binomial or $\chi^2$ statistics. Instead, they developed a new complex ad-hoc statistical method that shows random distribution in galaxy spin directions, and specifically in data from HSC. Source code for the method was also made available. The primary downside of the new method is that it is not able to identify asymmetry in the distribution of galaxy spin directions. Even when the new method is provided with synthetic data with extreme and obvious asymmetry, it still reports a null-hypothesis Universe with random distribution. That shows empirically that the method cannot sense asymmetry in the distribution of the directions of rotation of galaxies. While this further concludes that the distribution of galaxy spin direction as observed from Earth is not symmetric, it is not necessarily an indication of an anomaly in the large-scale structure of the Universe. The excessive number of galaxies that rotate in the opposite direction relative to the Milky Way can also be driven by the internal structure of galaxies and the physics of galaxy rotation. The phenomenon can be related directly to other puzzling anomalies such the Ho tension. Data are publicly available, and code is not needed to reproduce the results since only conventional statistics is used.
\end{abstract}


\section{Introduction}
\label{introduction}

The large-scale asymmetry in the distribution of galaxy spin directions has been suggested as early as the 1980's \citep{macgillivray1985anisotropy}. Using the unprecedented data collection power of digital sky surveys, that asymmetry has been observed by all relevant sky surveys such as SDSS \citep{longo2011detection,shamir2012handedness,shamir2020patterns,mcadam2023reanalysis}, PanSTARRS \citep{shamir2020patterns}, DES \citep{shamir2022asymmetry}, HST \citep{shamir2020pasa}, HCS \citep{shamir2024hcs}, DECam \citep{shamir2021large}, and DESI Legacy Survey \citep{shamir2022analysis}. The largest experiment included $\sim1.3\cdot10^6$ galaxies from the DESI Legacy Survey, allowing to perform binomial distribution analysis of the entire sky \citep{shamir2022analysis}.

The asymmetry exhibits itself in the form of a dipole axis that peaks at close proximity to the Galactic pole \citep{shamir2022analysis,sym15061190,sym15091704} and becomes stronger as the redshift gets higher \citep{shamir2020patterns,shamir2022large}, leading to the possibility that a higher number of galaxies that rotate in the opposite direction relative to the Milky Way is observed from Earth \citep{shamir2022analysis,sym15061190,sym15091704,shamir2024hcs}. Indeed, JWST has shown clearly that the number of galaxies that rotate in the opposite direction relative to the MW is far higher than the number of galaxies that rotate in the opposite direction relative to the MW \citep{Shamir_2024}.  


Recently, \cite{petal1} proposed a new statistical method that shows that the distribution of the galaxy spin directions is random, and use annotated HSC galaxies to make that claim \cite{petal1}. Instead of using standard binomial statistics or simple $\chi^2$ statistics, the new method is a complex ad-hoc method. This paper examines the new method empirically to test its ability to identify asymmetry in datasets of galaxies annotated by their direction of rotation.

\section{Testing the \citep{petal1} method with extremely asymmetric data}
\label{testing}

An interesting dataset that can be used to test the ability of the \cite{petal1} method is the dataset of mirrored galaxy images annotated by the {\it SpArcFiRe} method. {\it SpArcFiRe} was originally used in \citep{hayes2017nature}, and the experiment was reproduced in \citep{mcadam2023reanalysis}. As stated in the appendix of \citep{hayes2017nature}, {\it SpArcFiRe} has a consistent bias, which means that it produces an asymmetric dataset regardless of the physical properties of the Universe. Indeed, the dataset includes 139,852 galaxies, such that 70,672 galaxies rotating counterclockwise, and 69,180 galaxies rotate clockwise. The two-tailed binomial distribution probability to have such distribution by chance is $\sim0.00006$. The dataset is described in \citep{mcadam2023reanalysis}, and available for download at \url{https://people.cs.ksu.edu/~lshamir/data/sparcfire/}. Due to its asymmetry, that dataset was not used to claim for asymmetry in the distribution of galaxy spin directions, as it is difficult to separate the asymmetry driven by the algorithmic bias from the possible asymmetry driven by the real distribution of the spin directions of galaxies.

Despite the extreme asymmetry of the distribution in that dataset, the method used by \cite{petal1} does not identify any asymmetry in that data. As reported in the fourth row of Table 4 in \cite{petal}, the new method provides a p-value of 0.25. The dataset is notated as ``GAN M'' in \cite{petal}, and is the exact same dataset of mirrored galaxies taken from \url{https://people.cs.ksu.edu/~lshamir/data/sparcfire/}. That means that despite the extreme asymmetry of galaxy spin directions in that specific dataset, the new method of \cite{petal1} shows a null-hypothesis universe with no statistically significant asymmetry in the distribution of galaxy spins.  The fact that even in an extremely asymmetric dataset the new method is not able to show a statistically significant p value indicates that the new statistical method is not necessarily able to detect asymmetry in the distribution of galaxy spin directions.

The galaxies in the dataset used in \citep{shamir2024hcs} are from both ends of the Galactic pole, and therefore the asymmetry in one hemisphere offsets the asymmetry in the opposite hemisphere. To make the dataset clearly asymmetric, 150 galaxies that rotate clockwise were randomly assigned as galaxies that rotate counterclockwise. That led to a dataset with 6,628 clockwise galaxies and 6,899 counterclockwise galaxies, and therefore a dataset with extreme asymmetry. But when applying the method of \citep{petal1}, the method did not detect statistical significance. Table~\ref{datasets1} summarizes the asymmetry in the two experiments using two datasets with artificial asymmetry that is extremely high.

\begin{table}[h]
\scriptsize
\begin{tabular}{lcccc}
\hline
Dataset & \# cw  & \# ccw          & P               & P              \\
        &        &                 & (one-tailed)    &  (two-tailed)   \\
\hline
SDSS \citep{mcadam2023reanalysis}     & 69,180 &  70,672         &   0.00003 & 0.00006  \\
HCS \citep{shamir2024hcs}                   &  6,628  &   6,899         &    0.0029   &  0.0058   \\       
\hline
\end{tabular}
\caption{The number of galaxies rotating clockwise and counterclockwise in two datasets with artificial bias leading to extreme asymmetry. While the asymmetry in both cases is extreme, the method of \citep{petal1} indicates that the asymmetry is not statistically significant. The {\it p} values are based on standard binomial distribution such as the probability of a galaxy to rotate in a certain direction is 0.5. No code is needed for this conventional statistical analysis.}
\label{datasets1}
\end{table}



\section{Separation to galaxy rotation relative to the direction of rotation of the Milky Way}
\label{galactic_pole}

As shown in \citep{shamir2022analysis,sym15061190,sym15091704}, the asymmetry in the distribution of the spin directions of spiral galaxies is aligned with the Galactic pole. That was done using a very large dataset of $\sim1.3\cdot10^6$ galaxies analyzed through very simple binomial distribution. That experiment is also aligned with several other previous experiments that show that the dipole axis formed by the asymmetry in the distribution of galaxy spin directions is in close proximity to the Galactic pole. For instance, a summary of the results of all of these studies is displayed by Figure 7 in \citep{sym15091704}. 

According to these experiments, the number of galaxies rotating counterclockwise is higher in the hemisphere of the Northern Galactic pole, and lower in the hemisphere of the Southern Galactic pole. In simpler words, when observing galaxies from Earth, the number of galaxies that rotate in the opposite direction relative to the Milky Way is higher than the number of galaxies that rotate in the same direction relative to the Galactic pole. That means that the asymmetry in galaxy spin directions as observed from Earth might not be a feature of the large-scale structure of the Universe, but a feature of the internal structure of galaxies according which the rotational velocity of galaxies has a subtle but consistent effect on the light \citep{shamir2016asymmetry,shamir2017colour,shamir2020asymmetry,sym15061190}.

Table~\ref{pole_table} shows the same datasets discussed in Section~\ref{testing}, separated into galaxies that rotate in the same direction relative to the Milky Way and in the opposite direction relative to the Milky Way. In this case, the HCS dataset is used without adding artificial bias. That dataset can be accessed at \url{https://people.cs.ksu.edu/~lshamir/data/asymmetry_hsc/}. As the table shows, the datasets show a statistically significant higher number of galaxies that rotate in the opposite direction relative to the Milky Way. Once again, the method of \citep{petal1} was not able to identify any asymmetry, and report a null-hyphothesis Universe with no asymmetry in the distribution of the spin directions of galaxies.

\begin{table}[h]
\scriptsize
\begin{tabular}{lccc}
\hline
Dataset & \# MW  & \# OMW          & P                  \\
           &              &                       & (one-tailed)     \\
\hline
SDSS \citep{mcadam2023reanalysis}     & 69,272 &  70,580         &   0.00023   \\
HCS \citep{shamir2024hcs}                  &   6,630   &    6,847          &   0.031              \\       
\hline
\end{tabular}
\caption{The number of galaxies rotating in the same direction relative to the Milky Way (MW) and in the opposite direction relative to the Milky Way (OMW). While the asymmetry in both cases is statistically significant, the method of \citep{petal1} does not show any statistically significant asymmetry. The {\it p} values are determined by simple binomial distribution. The one-tailed p value is used since the higher number of galaxies that rotate in the opposite direction relative to the Milky Way has been shown in previous experiments \citep{shamir2022analysis,sym15061190,sym15091704}, and is therefore expected.}
\label{pole_table}
\end{table}



\section{Conclusion}
\label{conclusion}

Binomial distribution was introduced in the 18th century by Daniel Bernouli, and expanded by Ronald Fisher during the 20th century \citep{math10152680}. Since then, it has been used as an elemental form of statistics. In the sense of the distribution of the spin directions of spiral galaxies, binomial distribution can be used in its most basic form to test whether the number of galaxies that rotate in one direction is statistically different from the number of galaxies that rotate in the opposite direction, assuming 0.5 chance of a galaxy to rotate in one direction over the other. Results from all relevant digital sky surveys show consistent asymmetry in the distribution of galaxy spin directions. Specifically, they show that more galaxies as observed from Earth seem to rotate in the opposite direction relative to the Milky Way, compared to galaxies that rotate in the same direction relative to the Milky Way. That was done with smaller datasets as shown here, or with far larger datasets, allowing to apply binomial distribution in different parts of the sky to show an alignment between the asymmetry in galaxy spin directions and the Galactic pole \citep{shamir2022analysis}.

\begin{figure*}[h]
\centering
\includegraphics[scale=0.2]{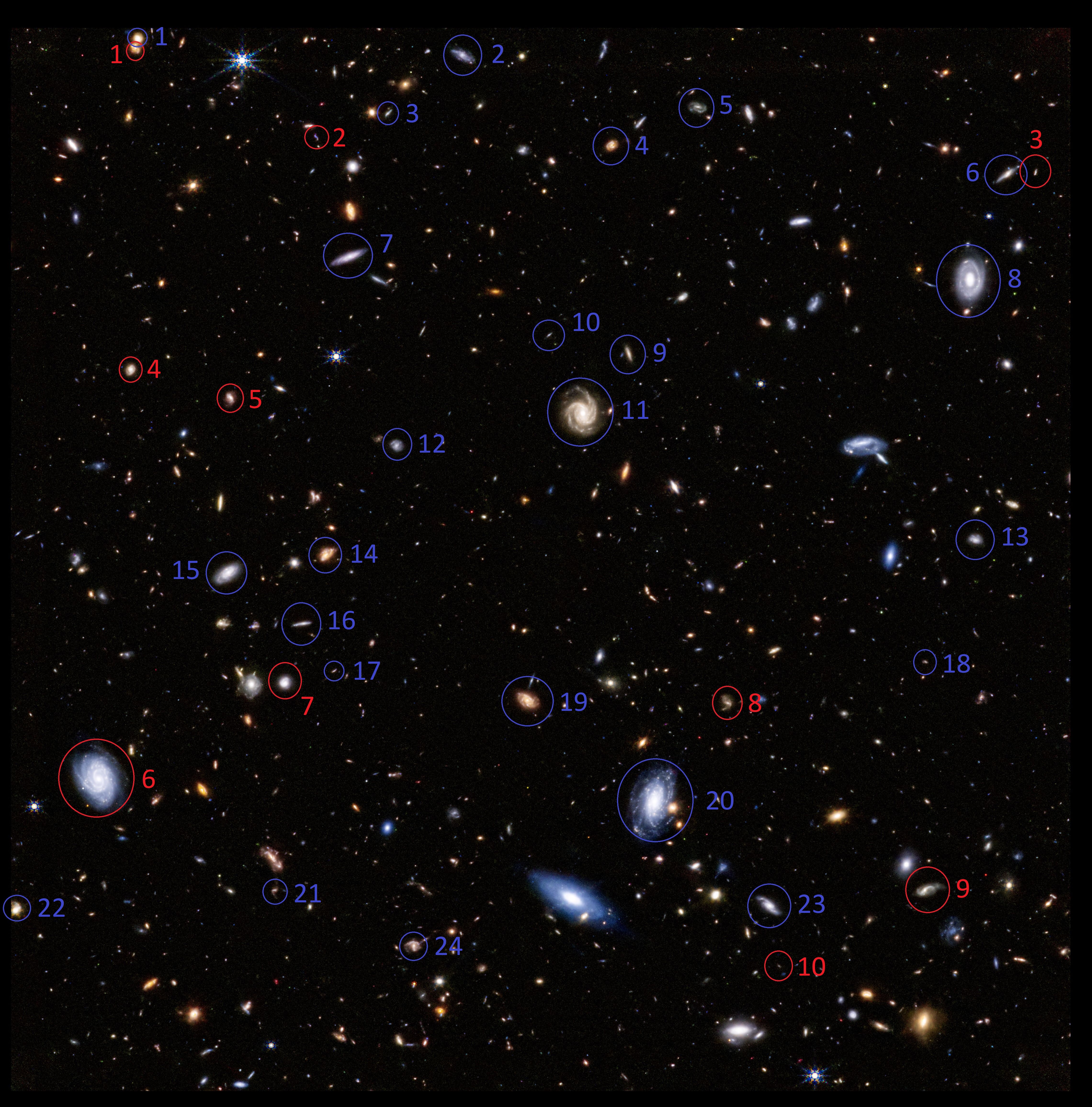}
\caption{Analysis of galaxies that rotate in opposite direction in JWST deep field taken in relatively close proximity to the Galactic pole \citep{Shamir_2024}. Simple binomial distribution shows 23 galaxies that rotate in the opposite direction relative to the Milky Way (blue), and just 10 galaxies that rotate in the same direction relative to the Milky Way (red). The binomial distribution probability to have such distribution by chance is $\sim0.012$ \citep{Shamir_2024}.}
\label{jwst}
\end{figure*}

A simple example is the JWST Advanced Deep Extragalactic Survey (JADES) deep field taken in close proximity to the Southern Galactic Pole \citep{Shamir_2024}. Figure~\ref{jwst} shows the annotation of the galaxies on top of the deep field image as taken from \citep{Shamir_2024}. Due to the relatively small number of galaxies, the image can be inspected manually. The deep field image shows 23 spiral galaxies that rotate in the opposite direction relative to the Milky Way, and merely 10 galaxies that rotate in the same direction relative to the Milky Way. The probability of such distribution to occur by chance can be determined by binomial distribution, showing probability of $\sim0.012$.

\cite{petal1} propose a new complex statistical method. But as shown empirically, including with the results published in \cite{petal}, the new method cannot identify asymmetry in galaxy spin directions even when the asymmetry is clear, including cases where artificial bias is added to the data to create an extremely asymmetric datasets. \cite{petal} also argued that the reproduction of previous results of \citep{longo2011detection} and \citep{mcadam2023reanalysis} provided different results than the results stated in these papers, as stated in Section 4.3 in \citep{petal}. That claim has also shown to be incorrect, with code and step-by-step instructions to easily reproduce the results of both papers. The full open code and step-by-step instructions can be found at \url{https://people.cs.ksu.edu/~lshamir/data/patel_desmond/}.

While numerous probes have shown cosmological-scale anisotropy that is not expected based on the standard model \citep{Aluri_2023}, the excessive number of galaxies that rotate in the opposite direction relative to the Milky Way may not necessarily be a feature of the large-scale structure of the Universe. The rotational velocity of the Milky Way relative to the observed galaxies can lead to subtle yet consistent photometric differences \citep{shamir2016asymmetry,shamir2017colour}, that also include brightness differences \citep{shamir2020asymmetry,sym15061190}. That can consequently lead to a higher number of galaxies detected from Earth based on the direction of their rotational velocity. Such effect can be explain other cosmological-scale anisotropies observed from Earth \citep{Aluri_2023}, and can also be related to anomalies such as the Ho tension, as was shown in \citep{sym15061190} using the SH0ES Ia supernovae.

\section*{Data availability}

Annotated HSC data is available at \url{https://people.cs.ksu.edu/~lshamir/data/asymmetry_hsc/}. Galaxies annotated by the {\it SPARCFIRE} algorithm can be accessed at \url{https://people.cs.ksu.edu/~lshamir/data/sparcfire/}.

\bibliographystyle{apalike}
\bibliography{main}

\end{document}